\documentclass[aps,prl,groupedaddress,twocolumn,nofootinbib]{revtex4-2}

\usepackage{amsmath}
\usepackage{amssymb}
\usepackage{graphicx}
\usepackage{color}
\usepackage{bm}
\usepackage{physics}
\usepackage{xr}
\usepackage{multirow}
\usepackage{latexsym}
\usepackage{comment}

\begin{document}

\title{Diffusion-driven self-assembly of emerin nanodomains at the nuclear envelope}

\author{Carlos D. Alas$^{1,2}$}
\author{Liying Wu$^{3}$}
\author{Fabien Pinaud$^{1,3,4}$}
\email{pinaud@usc.edu}
\author{Christoph A. Haselwandter$^{1,2}$}
\email{cah77@usc.edu}
\affiliation{$^1$Department of Physics and Astronomy, $^2$Department of Quantitative and Computational Biology, $^3$Department of Biological Sciences, and $^4$Department of Chemistry, University of Southern California, Los Angeles, CA 90089, USA}
\date{\today}

\begin{abstract}
Emerin, a nuclear membrane protein with important biological roles in mechanotransduction and nuclear shape adaptation, self-assembles into nanometer-size domains at the inner nuclear membrane. The size and emerin occupancy of these nanodomains change with applied mechanical stress as well as under emerin mutations associated with Emery-Dreifuss muscular dystrophy (EDMD). Through a combination of theory and experiment we show here that a simple reaction-diffusion model explains the self-assembly of emerin nanodomains. Our model yields quantitative agreement with experimental observations on the size and occupancy of emerin nanodomains for wild-type emerin and EDMD-associated mutations of emerin, with and without applied forces, and allows successful prediction of emerin diffusion coefficients from observations on the overall properties of emerin nanodomains. Our results provide a physical understanding of EDMD-associated defects in emerin organization in terms of changes in key reaction and diffusion properties of emerin and its nuclear binding partners.
\end{abstract}


\maketitle

Emerin is a largely disordered protein predominantly located at the nuclear envelope (NE) in mammalian cells and in cells of various other eukaryotic organisms \cite{berk13,koch14}. It is a major contributor to the maintenance of nuclear mechanics, as it participates in the transduction of mechanical signals across the nucleus double membrane \cite{lammerding05,rowat06}. Emerin mostly resides in the inner nuclear membrane (INM), where it interacts with multiple nuclear binding partners (NBPs) and NE components of the Linker of Nucleoskeleton and Cytoskeleton (LINC) complexes, to transfer on the nuclear matrix forces that are generated by the cytoskeleton and that travel through the nucleus's outer and inner membranes via LINC complexes \cite{Fernandez2022,mejat10,kim15,crisp06,lombardi11,chang15}. Mutations in emerin that impact its interactions with NBPs and its self-assembly into nanodomains, such as $\Delta95$-$99$, Q133H, or P183H mutations \cite{berk13,Fernandez2022,berk14,herrada15,samson17}, correlate with abnormal responses of the NE to mechanical stress \cite{Fernandez2022}. In cells exposed to extensive forces, such as skeletal and cardiac cells, these aberrant responses result in Emery-Dreifuss muscular dystrophy (EDMD) \cite{emery66}.

Numerous studies have highlighted the significance of emerin’s disordered region for its biological function in nuclear mechanics \cite{Fernandez2022,berk14,herrada15,samson17}. This region mediates emerin self-assembly as well as binding to NBPs that regulate the nuclear architecture, including lamins, nuclear actin, LINC complex proteins and other molecular partners \cite{berk13,haque10}. The expected structural flexibility of its disordered region likely allows emerin to adopt various conformations that, in turn, modulate emerin's self-assembly and its engagement with different NBPs at the INM. Biochemistry studies have indeed indicated that wild-type (WT) emerin is associated with two different nucleoskeletal neighborhoods at the NE \cite{berk13b} and recent single-molecule imaging studies have revealed that, at steady state, it distributes into rapidly- and slowly-diffusing emerin populations, the latter forming stable INM nanodomains characterized by elevated emerin concentrations \cite{Fernandez2022}. Those imaging studies also showed that adequate nuclear responses to mechanical challenges induced by cell micropatterning require controlled changes in the diffusion properties and spatial organization of both types of emerin complexes \cite{Fernandez2022}. In effect, compared to WT emerin, the aforementioned EDMD-inducing emerin mutants display either insufficient or excessive self-assembly into nanodomains, both of which result in defective nuclear shape adaptations against force \cite{Fernandez2022}. Modulation of the self-assembly of emerin into INM nanodomains is therefore a central determinant of NE response to forces as it prevents deleterious nucleus deformations typically observed in EDMD.

The spatial patterns of emerin observed at the INM, the distinction between slowly- and rapidly-diffusing emerin complexes, and the observed dependence of emerin nanodomains on emerin diffusion are reminiscent of molecular patterns resulting from reaction-diffusion processes \cite{Turing1952,Epstein1998,bailles22}. A slowly-diffusing particle species thereby interacts with other molecules so as to locally activate increased molecule concentrations, while a rapidly-diffusing particle species inhibits increased molecule concentrations. In this Letter we combine theory and experiment to explain the self-assembly of emerin nanodomains in terms of such a reaction-diffusion model of emerin complexes. After validating our model for WT emerin under no mechanical stress, we employ our model to understand how and why WT emerin nanodomains respond to force application. We then use our model to establish how defects in nanoscale organization for EDMD-associated emerin mutants correlate with changes in key reaction and diffusion properties of emerin and its NBPs. Our results suggest that the self-assembly and plasticity of emerin nanodomains result from the interaction of slowly-diffusing emerin complexes that can locally bind other emerin, and rapidly-diffusing emerin complexes that inhibit increased molecule concentrations through steric constraints.

\textit{Modeling emerin nanodomains.}---Superresolution microscopy experiments on emerin nanodomains report the diffusion coefficients of rapidly- and slowly-diffusing emerin populations at the INM, the size of emerin nanodomains, and their molecular density along the INM \cite{Fernandez2022, SM}. The question thus arises to what extent the observed emerin densities along the INM can be accounted for based on the measured emerin diffusion coefficients together with the observed nanodomain sizes. We have addressed this question through a simple model of emerin diffusion in heterogeneous media \cite{Li2017Sep,SM}. We find that this model yields, with no adjustable parameters, the observed localization of WT emerin to nanodomains, without the need to invoke cellular structures that confine emerin to particular membrane regions. These results suggest that the observed distributions of emerin along the INM can be understood quantitatively from emerin’s diffusion properties, which we take as our starting point for modeling the self-assembly of emerin nanodomains.

Slowly-diffusing emerin is thought to interact with NBPs so as to facilitate binding to other emerin, while rapidly-diffusing emerin also interacts with NBPs but is not thought to produce higher-order emerin structures \cite{Fernandez2022}. In our model of emerin nanodomain self-assembly we therefore allow for slowly- and rapidly-diffusing emerin-NBP complexes at the INM to have distinct emerin and NBP binding properties. We denote these two types of emerin-NBP complexes by $A$ and $I$ with diffusion coefficients $\nu_A$ and $\nu_I>\nu_A$, respectively. We assume that the slowly-diffusing $A$ complexes can transiently bind other emerin or NBPs to locally increase the concentration of $A$ and $I$. In contrast, we assume that the rapidly-diffusing $I$ complexes do not bind other emerin or NBPs but can crowd the INM. Thus, $A$ complexes locally \textit{activate} increased concentrations of $A$ and $I$, while $I$ complexes \textit{inhibit} increased concentrations of $A$ and $I$ through steric constraints, such that the formation of $A$ and $I$ is significantly enhanced in membrane regions with elevated concentrations of $A$ complexes. To see how these reaction-diffusion properties of emerin can yield nanodomain self-assembly via a Turing mechanism \cite{Turing1952,Haselwandter2015}, consider a random distribution of $I$ and $A$ complexes along the INM. If, at some INM location, there is a local excess of $A$ over $I$, then $A$ will tend to locally increase the concentrations of both $A$ and $I$. Since $I$ complexes diffuse away more rapidly, this produces a positive feedback elevating the concentration of emerin molecules at that INM location. Eventually, a steady state is reached when enough $I$ complexes are drawn in to balance the local population of $A$ complexes, producing a stable pattern of emerin nanodomains.

To quantify the above mechanism for emerin nanodomain self-assembly it is necessary to specify reactions for $I$ and $A$ complexes, for which we employ experiments on WT emerin \cite{Fernandez2022}. While we show here that the emerin-NBP interactions captured by $I$ and $A$ complexes at the INM are sufficient to produce the observed emerin nanodomains, we also note that the rapidly- and slowly-diffusing emerin populations seen in experiments most likely encompass more than just these two types of molecular complexes, which a more detailed model would take into account. Based on observations that, before it accumulates at the INM, emerin also distributes in the endoplasmic reticulum and outer nuclear membranes where no NBP are present, we assume that $I$ and $A$ complexes in the INM can both assemble from or dissociate into a pool of emerin and NBPs that lack the molecular requirements to form $I$ or $A$ complexes. For simplicity, we take the spontaneous assembly of $I$ and $A$ complexes participating in emerin nanodomain formation to be negligible compared to their spontaneous disassembly, $I\xrightarrow{f_1}\varnothing$ and $A\xrightarrow{g_1}\varnothing$ with disassociation rates $f_1$ and $g_1$, a model assumption that can easily be lifted \cite{SM}.

When imaging WT emerin, rapidly-diffusing emerin are primarily observed outside emerin nanodomains of diameter $22\pm11$~nm while slowly-diffusing emerin are primarily found inside nanodomains \cite{Fernandez2022}. We thus assume that $I$ complexes can diffuse over a length scale of at least $\sim 33/2$~nm over their lifetime at the INM, so that they can readily diffuse out of emerin nanodomains, while, consistent with $A$ complexes being nearly immobile at the INM \cite{Fernandez2022}, $A$ complexes stay localized to a molecular length scale, which we set at $\sim 1$~nm. From the root-mean-square displacements $2 \sqrt{\nu_{I}/f_1}$ and $2 \sqrt{\nu_{A}/g_1}$ we thus estimate $f_1\approx30~\mathrm{s}^{-1}$ and $g_1\approx40 f_1$ for the diffusion coefficients $\nu_I\approx2\times10^{-3}\mu\mathrm{m}^2/\mathrm{s}$ and $\nu_A \approx 3 \times10^{-4} \mu \mathrm{m}^2/\mathrm{s}$ measured for WT emerin \cite{Fernandez2022}. These values of $f_1$ and $g_1$ can be changed by $>50\%$ in our model to obtain WT emerin nanodomains  with similar properties.

At the most basic level, single $A$ complexes may produce local increases in emerin concentration at the INM by facilitating the formation of $I$ and $A$ complexes, $A + \varnothing \xrightarrow{f_2} A + I$ and $A + \varnothing \xrightarrow{g_2} 2 A$. Furthermore, the experimental phenomenology of emerin nanodomains suggests that $A$ complexes can form higher-order oligomers \cite{Fernandez2022}. We therefore allow for a higher-order reaction in which two $A$ complexes facilitate the formation of another $A$ complex, $2 A + \varnothing \xrightarrow{g_3} 3 A$. We estimate the values of the reaction rates $f_2$, $g_2$, and $g_3$ from $f_1$ and $g_1$. In particular, experiments on WT emerin indicate that nanodomains are predominantly composed of $A$, rather than $I$ complexes \cite{Fernandez2022}, suggesting that $f_2 < f_1$. We set here $f_2=f_1/2$ for WT emerin. Due to the slow diffusion of $A$ complexes, the leading-order dissociation and assembly rates of $A$ complexes must be approximately equal to each other so that a non-trivial steady state can be achieved, and we thus set $g_2 = g_1$ for WT emerin. Finally, we assume that, as specified mathematically below, higher-order reactions have a smaller propensity to occur than lower-order reactions. We therefore set $g_3=g_2/10$ for WT emerin. Other choices for the values of $f_2$, $g_2$, and $g_3$ give similar results for WT emerin provided that $f_2 \lesssim f_1$, $g_1 \approx g_2$, and $g_3 \ll g_2$.

We quantify the fractional area coverage of $I$ and $A$ complexes at a particular INM location $(x,y)$ and time $t$ by the fields $I(x,y,t)$ and $A(x,y,t)$ with $0\leq I \leq 1$ and $0\leq A \leq 1$, where the upper bounds on $I$ and $A$ account for steric constraints. We rescale the rates of all reaction and diffusion processes locally increasing $I$ or $A$ by a steric factor $S=1-I-A$ so as to ensure that $0 \leq I + A \leq 1$. At the mean-field level, the fields $I$ and $A$ are then governed by the reaction-diffusion equations~\cite{satulovsky96,lugo08,Haselwandter2015,Li2017May}
\begin{eqnarray}\label{eqIflow}
	\frac{\partial I}{\partial t} &=& \mathrm{F}\left(I,A\right)+\nu_I \nabla \cdot \left[\left(1-A\right)\nabla I+I\nabla A\right]\,,\\ \label{eqAflow}
	\frac{\partial A}{\partial t} &=&\mathrm{G}\left(I,A\right)+\nu_A \nabla \cdot \left[\left(1-I\right)\nabla A+A\nabla I\right]\,,
\end{eqnarray}
where the polynomials $\mathrm{F}$ and $\mathrm{G}$ describe the aforementioned reaction dynamics of $I$ and $A$ complexes,
\begin{eqnarray}\label{polyF}
	\mathrm{F}\left(I,A\right) &=& -f_1 I + f_2 S A\,,\\ \label{polyG}
	\mathrm{G}\left(I,A\right) &=& -g_1 A + g_2 S A + \frac{g_3}{2 \bar A} S A^2\,,
\end{eqnarray}
and we denote the values of $I$ and $A$ at the homogeneous steady state $\mathrm{F} = \mathrm{G} = 0$ by $\bar I$ and $\bar A$, respectively. The factor $1/2$ in the last term in Eq.~(\ref{polyG}) arises because this term describes a second-order reaction involving two (indistinguishable) $A$ complexes \cite{Li2017May}. Furthermore, we rescale $g_3$ in Eq.~(\ref{polyG}) by the characteristic value $A=\bar A$ so as to permit direct numerical comparisons of $g_2$ and $g_3$, which allows us to fix $g_3$ in terms of $g_2$ so that $g_3 \ll g_2$ even though these two parameters are associated with reactions of different order. Note that $\bar I$ and $\bar A$ depend on all reaction rates in the model \cite{SM}. Together with Eqs.~(\ref{polyF}) and~(\ref{polyG}), the reasoning above fixes all parameter values in Eqs.~(\ref{eqIflow}) and~(\ref{eqAflow}) for WT emerin. To study the self-assembly of emerin nanodomains we numerically solve Eqs.~(\ref{eqIflow}) and~(\ref{eqAflow}) starting from random initial conditions about $\left(I,A\right)=\left(\bar I, \bar A\right)$~\cite{SM}.

\begin{figure}[t!]
	\includegraphics{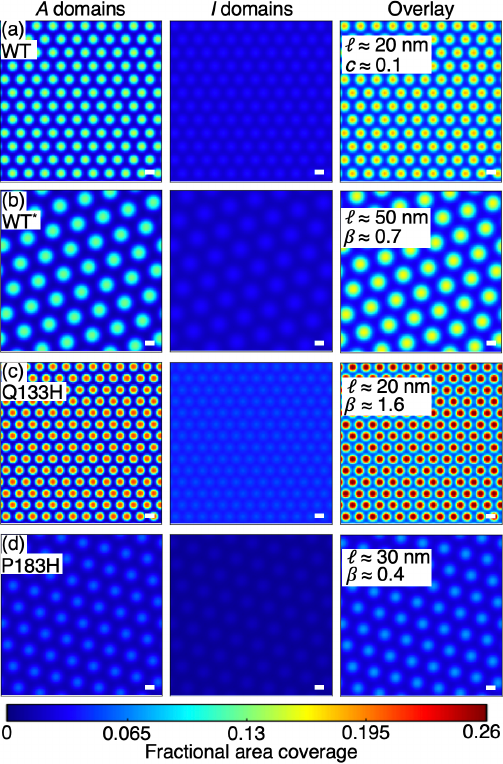}
	\centering
	\caption{Numerical solutions of Eqs.~(\ref{eqIflow}) and~(\ref{eqAflow}) with Eqs.~(\ref{polyF}) and~(\ref{polyG}) for (a) WT emerin, (b) WT emerin under force (WT$^\star$), (c) Q133H emerin, and (d) P183H emerin. All solutions correspond to steady states of Eqs.~(\ref{eqIflow}) and~(\ref{eqAflow}) with Eqs.~(\ref{polyF}) and~(\ref{polyG}) obtained from random initial conditions at $t=0$. The left, middle, and right density maps show $A(x,y,t)$, $I(x,y,t)$, and $A(x,y,t) + I(x,y,t)$ at $t=100 \tau$, respectively, where the characteristic time scale $\tau$ follows from a linear stability analysis and is given by (a) $\tau\approx6~\mathrm{s}$, (b) $\tau\approx17~\mathrm{s}$, (c) $\tau\approx3~\mathrm{s}$, and (d) $\tau\approx32~\mathrm{s}$ \cite{SM}. The values of the reaction and diffusion parameters in Eqs.~(\ref{eqIflow}) and~(\ref{eqAflow}) with Eqs.~(\ref{polyF}) and~(\ref{polyG}) were chosen as explained in the main text. Scale bars, 20~nm.}
	\label{fig:1}
\end{figure}

\textit{Organization of WT emerin.}---We validate our reaction-diffusion model of emerin nanodomain self-assembly based on experimentally measured diffusion coefficients $\nu_I$ and $\nu_A$ of WT emerin at the INM \cite{Fernandez2022}. For the parameter values described above, Eqs.~(\ref{eqIflow}) and~(\ref{eqAflow}) yield, starting from random initial conditions, spontaneous self-assembly of emerin nanodomains [see Fig.~\ref{fig:1}(a)]. We quantify the size of these nanodomains through a linear stability analysis to calculate the characteristic nanodomain diameter $\ell$ implied by Eqs.~(\ref{eqIflow}) and~(\ref{eqAflow}) \cite{SM}. We find $\ell\approx20$~nm in the steady-state of the system, which agrees with the nanodomain diameter measured for WT emerin, $\ell=22\pm11$~nm \cite{Fernandez2022}, and the numerical solutions in Fig.~\ref{fig:1}(a). In agreement with experimental observations we find that the emerin populations in WT nanodomains are dominated by the slowly-diffusing $A$ complexes rather than the rapidly-diffusing $I$ complexes [Fig.~\ref{fig:1}(a)]. We obtained results similar to those shown in Fig.~\ref{fig:1}(a) when we allowed for additional steric effects arising from a pool of emerin (or other NBPs) not accounted for through $I$ and $A$ complexes~\cite{SM}.

We establish further links between our model results and experimental observations through the average fraction of the nanodomain area covered by $I$ and $A$ complexes, which we denote by $c$. To estimate $c$ from numerical solutions of Eqs.~(\ref{eqIflow}) and~(\ref{eqAflow}) we first find, for a given nanodomain, the grid point associated with the maximum of $\left(I+A\right)$ in the steady state of the system. We then average $\left(I+A\right)$ over all grid points within a radius $\ell/2$, rounded to the nearest multiple of the grid spacing, about this (local) maximum of $\left(I+A\right)$. We carry out this procedure for five nanodomains and average the results to obtain $c$. This last step was, strictly speaking, not necessary, since $c$ evaluated over a single nanodomain and $c$ evaluated over multiple nanodomains yield similar results. We estimate $c$ from experiments based on the measured emerin numbers in nanodomains, the measured diameter of nanodomains, and the INM area occupied by emerin ($\approx 1$~nm$^2$--4~nm$^2$) \cite{SM}. We find $c\approx0.1$ from Eqs.~(\ref{eqIflow}) and~(\ref{eqAflow}), with experiments on WT emerin giving the values $c\approx0.03$--$0.1$~\cite{SM}.

The above results show that Eqs.~(\ref{eqIflow}) and~(\ref{eqAflow}) yield a WT emerin nanodomain diameter $\ell$ and fractional area coverage $c$ close to experimental estimates. However, a few notable discrepancies between experiment and theory deserve comment. First, we note that our model produces closely spaced nanodomains with a uniform size and shape, while experiments show nanodomains with irregular sizes and shapes that tend to be more widely spaced than the nanodomains in Fig.~\ref{fig:1}(a) \cite{Fernandez2022,SM}. This discrepancy arises, on the one hand, from the mean-field character of Eqs.~(\ref{eqIflow}) and~(\ref{eqAflow}), which neglect the intrinsic noise associated with the reaction and diffusion processes considered here. Such noise can produce irregular domain shapes, sizes, and spacings, and even result in nanodomain linkage \cite{Law2021}. On the other hand, we note that the INM contains large membrane structures, such as nuclear pore complexes, that restrict the membrane area available for emerin nanodomain self-assembly and, thus, increase the effective separation of emerin nanodomains seen in experiments. Furthermore, the value of $c$ predicted from Eqs.~(\ref{eqIflow}) and~(\ref{eqAflow}) is at the upper bound of the range of values of $c$ suggested by experiments. This discrepancy likely arises because, due to a lack of detailed experimental data on how emerin interacts with NBPs in nanodomains, we only consider the size of emerin when estimating $c$ from experiments and thus effectively neglect the finite size of NBPs \cite{SM}. As a result, our experimental estimates of $c$ likely underestimate $c$. Similar considerations apply to the scenarios we consider next.

\textit{Organization of WT emerin under force (WT$^\star$).}---Subjecting cells to nuclear mechanical stress using $10~\mu$m or $15~\mu$m wide micropatterns induces an increase in WT emerin nanodomain size by approximately three-fold, from $\ell = 22\pm11~\mathrm{nm}$ to $\ell = 60\pm13~\mathrm{nm}$ \cite{Fernandez2022}. Furthermore, force experiments indicate that for WT$^\star$ nanodomains the value of $c$ is a fraction $\beta \approx 0.6$ of the value of $c$ obtained for WT nanodomains. Experimental measurements further indicate that, possibly due to a mechanical stress-induced disruption in the interactions between emerin and NBPs, the diffusion coefficients $\nu_I$ and $\nu_A$ are approximately doubled for WT$^\star$ emerin as compared to WT emerin, with $\nu_I=4\times10^{-3}\mu\mathrm{m}^2/\mathrm{s}$ and $\nu_A=6\times10^{-4}\mu\mathrm{m}^2/\mathrm{s}$ \cite{Fernandez2022}.

Adjusting $\nu_I$ and $\nu_A$ in our model to account for WT$^\star$ emerin while using the same reaction dynamics as for WT emerin, we found that $\ell$ increased to $\ell\approx30$~nm with $\beta \approx 0.9$. This suggests that the observed changes in $\nu_I$ and $\nu_A$ can partially, but not fully, account for the observed changes in the size and density of WT emerin nanodomains under force. Considering the reduced experimental value of $\beta$ for WT$^\star$ nanodomains, we hypothesized that mechanical stress diminishes the relative strength of higher-order interactions, which facilitate the assembly of emerin complexes in our model. To test whether such a modification of the reaction dynamics can explain the observed changes in nanodomain size and density we decreased $g_3$ by $50\%$ relative to $g_2$, to $g_3 = g_2/20$ [see Fig.~\ref{fig:1}(b)]. This modification increased the nanodomain size to $\ell\approx50~\mathrm{nm}$ with $\beta\approx0.7$. Decreasing $g_3$ further by $10\%$ relative to $g_2$ resulted in $\ell = 60~\mathrm{nm}$ and $\beta\approx0.6$. Thus, the observed increases in the emerin diffusion coefficients together with a decrease in the relative strength of higher-order interactions seem to underlie the observed response of WT emerin nanodomains to mechanical stress.

\textit{Organization of Q133H emerin mutant.}---The Q133H mutation of emerin was observed to yield nanodomains of diameter $\ell = 19\pm12$~nm under no mechanical stress, which is statistically identical to the nanodomain size $\ell \approx 20$~nm associated with WT emerin, while the emerin density in Q133H nanodomains was increased by approximately $50\%$ compared to WT emerin, $\beta \approx 1.5$ \cite{Fernandez2022}. Furthermore, Q133H emerin was found to diffuse somewhat more rapidly than WT emerin, with the diffusion coefficients $\nu_I\approx 3\times10^{-3}$~$\mu\mathrm{m}^{2}/\mathrm{s}$ and $\nu_A \approx 4\times10^{-4}$~$\mu\mathrm{m}^{2}/\mathrm{s}$ \cite{Fernandez2022}. Adjusting $\nu_I$ and $\nu_A$ in our model to account for Q133H emerin but using the same reaction dynamics as for WT emerin, we found that $\ell$ increased by $10\%$ while $c$ remained approximately unchanged compared to WT emerin. Thus, the observed changes in Q133H emerin nanodomains seem to rely on changes in the emerin reaction properties.

It has been proposed that the Q133H mutation of emerin increases the potential of emerin-NBP complexes to bind additional emerin \cite{Fernandez2022}. We can quantify and test this hypothesis by noting that, in our model, $A$ complexes represent emerin-NBP complexes that can bind additional emerin. We therefore assume that the Q133H mutation of emerin leads to a more pronounced dependence of the reaction dynamics in Eqs.~(\ref{polyF}) and~(\ref{polyG}) on reactions driven by $A$ complexes, which we implemented through a uniform percentage increase in the strength of these reactions. Figure~\ref{fig:1}(c) shows model results obtained with an increase by $30\%$ in $f_2$, $g_1$, $g_2$, and $g_3$ compared to WT emerin. In agreement with experiments, we now find Q133H emerin nanodomains with a diameter $\ell\approx20~\mathrm{nm}$ and $\beta\approx1.6$. The agreement between model results and experiments suggests that the observed changes in Q133H nanodomains result from more rapid emerin diffusion together with an elevated propensity of emerin-NBP complexes to bind additional emerin.

\textit{Organization of P183H emerin mutant.}---The P183H mutation of emerin was observed to yield nanodomains with a diameter $\ell=35\pm12~\mathrm{nm}$ and an emerin density in nanodomains that was decreased by approximately $70\%$ compared to WT emerin, $\beta \approx 0.3$ \cite{Fernandez2022}. Furthermore, P183H emerin was observed to diffuse more slowly than WT emerin, with the diffusion coefficients $\nu_I\approx1\times10^{-3}$~$\mu\mathrm{m}^{2}/\mathrm{s}$ and $\nu_A\approx1\times10^{-4}$~$\mu\mathrm{m}^{2}/\mathrm{s}$ \cite{Fernandez2022}. Adjusting $\nu_I$ and $\nu_A$ in our WT model to account for P183H emerin but not changing any reaction rates, we find $\ell\approx10$~nm and $\beta\approx1.1$. Thus, similarly as for Q133H emerin, the observed changes in P183H emerin nanodomains seem to rely on changes in the emerin reaction properties.

The P183H mutation of emerin is thought to decrease the potential of emerin-NBP complexes to bind additional emerin \cite{Fernandez2022}. Decreasing, in analogy to Q133H, $f_2$, $g_1$, $g_2$, and $g_3$ by $30\%$ compared to WT emerin we find emerin nanodomains that were smaller and more dense than the nanodomains observed in experiments on P183H emerin, with $\ell\approx 20$~nm and $\beta\approx0.7$. We reasoned that, similarly as in the case of WT$^\star$ emerin, the P183H mutation may produce a decrease in the relative strength of higher-order interactions facilitating the assembly of emerin complexes. Figure~\ref{fig:1}(d) shows model results obtained for P183H emerin with, in analogy to Q133H and WT$^\star$ emerin, a decrease in $f_2$, $g_1$, and $g_2$ by $30\%$ but a decrease in $g_3$ by $60\%$ compared to WT emerin. In agreement with experiments, we find P183H nanodomains with a diameter $\ell\approx 30$~nm and $\beta\approx0.4$. Thus, the observed changes in P183H nanodomains appear to rely on a decreased propensity of emerin-NBP complexes to bind additional emerin, together with a decrease in the relative strength of higher-order interactions that facilitate the assembly of emerin complexes.

\begin{figure}[t!]
	\includegraphics{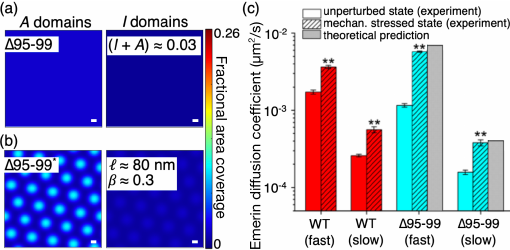}
	\centering
	\caption{Numerical solutions of Eqs.~(\ref{eqIflow}) and~(\ref{eqAflow}) with Eqs.~(\ref{polyF}) and~(\ref{polyG}) at $t=100 \tau$ as in Fig.~\ref{fig:1} but for (a) $\Delta95$-$99$ emerin with $\tau\approx~128$~s and (b) $\Delta95$-$99^\star$ emerin with $\tau\approx65$~s. Scale bars, 20~nm. (c) Measured and predicted diffusion coefficients of $\Delta95$-$99^\star$ emerin. The corresponding diffusion coefficients of WT and $\Delta95$-$99$ emerin are shown for comparison. T-test for experimental results on unperturbed versus mechanically stressed cells (**): $p<0.01$.
	}
	\label{fig:2}
\end{figure}

\textit{Organization of $\Delta95$-$99$ emerin mutant.}---In the absence of mechanical stress, the $\Delta95$-$99$ mutation of emerin was observed to yield an approximately random emerin distribution across the INM, with little-to-no nanodomain formation, and with diffusion coefficients $\nu_I\approx1\times10^{-3}$ $\mu\mathrm{m}^{2}/\mathrm{s}$ and $\nu_A\approx2\times10^{-4}$ $\mu\mathrm{m}^{2}/\mathrm{s}$ for the rapidly- and slowly-diffusing emerin populations \cite{Fernandez2022}. We hypothesized that, similarly to P183H emerin, the $\Delta95$-$99$ mutation of emerin decreases the potential of emerin-NBP complexes to bind additional emerin. Taking, for simplicity, the reaction dynamics of $\Delta95$-$99$ emerin to be identical to those of P183H emerin in Fig.~\ref{fig:1}(d), we find that Eqs.~(\ref{eqIflow}) and~(\ref{eqAflow}) yield, for the diffusion coefficients measured for $\Delta95$-$99$ emerin, homogeneous $I$ and $A$ distributions and no nanodomains [see Fig.~\ref{fig:2}(a)]. Similar results are obtained when the strength of the reactions in Eqs.~(\ref{polyF}) and~(\ref{polyG}) driven by $A$ complexes is decreased by as little as $20\%$ compared to WT emerin. Thus, we find that a decreased propensity of emerin-NBP complexes to bind additional emerin with, compared to P183H emerin, more rapid diffusion of $A$ complexes seem to underlie the failure of $\Delta95$-$99$ emerin to self-assemble into~nanodomains.

\textit{Organization of $\Delta95$-$99$ emerin under force ($\Delta95$-$99^\star$).}---Although diffusion coefficients for $\Delta95$-$99^\star$ emerin were not measured in previous experiments \cite{Fernandez2022}, it was observed that mechanical stress, induced by placing cells into $10~\mu$m micropatterns, yields $\Delta95$-$99^\star$ nanodomains with an increased diameter, $\ell = 75\pm20$~$\mathrm{nm}$, and a decreased emerin density, $\beta \approx 0.3$, compared to WT (and WT$^\star$) emerin \cite{Fernandez2022}. We combined these experimental observations with our model results to estimate the reaction and diffusion properties of $\Delta95$-$99^\star$ emerin. In particular, based on our results for WT$^\star$ emerin, we expect mechanical stress to weaken interactions between emerin and NBPs so as to decrease $g_3$ and increase $\nu_A$. In analogy to WT$^\star$ emerin, we therefore reduced $g_3$ by $50\%$ compared to $\Delta95$-$99$ emerin in Fig.~\ref{fig:2}(a) and increased $\nu_A$ by a factor of two, to $\nu_A\approx4\times10^{-4}$~$\mu\mathrm{m}^{2}/\mathrm{s}$. We hypothesized that the combined effects of the $\Delta95$-$99$ mutation and of mechanical stress largely decouple $I$ from NBPs, such that $\Delta95$-$99^\star$ emerin shows a more pronounced increase in $\nu_I$ than in $\nu_A$. Upon increasing $\nu_I$ from the value $\nu_I\approx1\times10^{-3}$ $\mu\mathrm{m}^{2}/\mathrm{s}$ measured for $\Delta95$-$99$ emerin in the absence of mechanical stress, we find that a seven-fold increase in $\nu_I$, to $\nu_I\approx7\times10^{-3}$~$\mu\mathrm{m}^{2}/\mathrm{s}$, yields self-assembly of $\Delta95$-$99^\star$ nanodomains similar to those found experimentally, with $\ell\approx80$~nm and $\beta\approx0.3$ [see~Fig.~\ref{fig:2}(b)].

To test the robustness of our model and assess whether our theoretical predictions of $\nu_A$ and $\nu_I$ for $\Delta95$-$99^\star$ emerin align with experiments, we proceeded to experimentally measure the diffusion coefficients of $\Delta95$-$99^\star$ emerin at the INM for cells mechanically stressed on $10~\mu$m micropatterns, as in Ref.~\cite{Fernandez2022}. We found slowly- and rapidly-diffusing $\Delta95$-$99^\star$ emerin populations with $\nu_A= 3.8\pm0.3\times10^{-4}$~$\mu\mathrm{m}^{2}/\mathrm{s}$ and $\nu_I= 5.8\pm0.1\times10^{-3}$~$\mu\mathrm{m}^{2}/\mathrm{s}$, respectively, similar to our theoretical predictions [see Fig.~\ref{fig:2}(c)]. Interestingly, Eqs.~(\ref{eqIflow}) and~(\ref{eqAflow}) do not yield emerin nanodomains for these specific values of $\nu_A$ and $\nu_I$ if, as for WT$^\star$ emerin, $g_3$ is reduced by $50\%$ compared to $\Delta95$-$99$ emerin in Fig.~\ref{fig:2}(a), but Eqs.~(\ref{eqIflow}) and~(\ref{eqAflow}) do yield $\Delta95$-$99^\star$ nanodomains if $g_3$ is reduced by (slightly) less than $50\%$. For instance, decreasing $g_3$ by $45\%$ compared to $\Delta95$-$99$ emerin in Fig.~\ref{fig:2}(a) while setting $\nu_A= 3.8\times10^{-4}$~$\mu\mathrm{m}^{2}/\mathrm{s}$ and $\nu_I= 5.8\times10^{-3}$~$\mu\mathrm{m}^{2}/\mathrm{s}$, Eqs.~(\ref{eqIflow}) and~(\ref{eqAflow}) yield emerin nanodomains similar to those in Fig.~\ref{fig:2}(b) and found experimentally for $\Delta95$-$99^\star$ emerin, with $\ell\approx80$~nm and $\beta\approx0.3$. These results suggest that the oligomerization of $\Delta95$-$99$ emerin is less sensitive to mechanical stress than that of WT emerin, which is consistent with the $\Delta95$-$99$ emerin mutant being unable to induce adequate compliance of the NE against mechanical stress \cite{Fernandez2022}. Thus, the combined effects of more rapid diffusion of $I$ and $A$ complexes, with a greater percentage increase in $\nu_I$ than in $\nu_A$, and a decrease in the relative strength of higher-order interactions seem to underlie the observed force-induced transition from a random distribution of $\Delta95$-$99$ emerin to self-assembled emerin nanodomains \cite{Fernandez2022}.

\textit{Conclusion.}---We have introduced here a simple physical model that provides a quantitative description of the formation of emerin nanodomains at the INM. The model predicts that the self-assembly of emerin nanodomains can result from a Turing mechanism in which emerin form slowly- or rapidly-diffusing complexes with NBPs that activate or inhibit locally increased emerin concentrations at the INM, respectively. Our model suggests that rapidly-diffusing emerin play a critical role in the self-assembly of stable emerin nanodomains, as initially implied by superresolution imaging experiments that allowed a quantitative characterization of slowly- and rapidly-diffusing emerin populations at the INM. On the one hand, we showed how the measured diffusion properties of emerin give rise to the observed supramolecular organization of emerin. On the other hand, our model establishes a connection between observed changes in the supramolecular organization, and associated biological roles, of emerin and modifications in key molecular properties of emerin. In particular, we identified key changes in the reaction and diffusion properties of emerin underlying the observed alterations of emerin nanodomains under mechanical stress and EDMD-associated mutations of emerin. The physical model described here provides new avenues for the control of nanodomain self-assembly for such mutated forms of emerin through modification of emerin reaction or diffusion properties.

\textit{Acknowledgments.}---This work was supported by the National Science Foundation through Award No.~MCB-2202087 (to F.P. and C.A.H.) and through Award No.~DMR-2051681 (to C.A.H.). We thank C.~M. Doucet and E. Margeat for helpful discussions.

\bibliographystyle{apsrev4-2}

\bibliography{references}

\end{document}


\title{Diffusion-driven self-assembly of emerin nanodomains at the nuclear envelope\\\textit{Supplemental Material}}

\author{Carlos D. Alas$^{1,2}$}
\author{Liying Wu$^{3}$}
\author{Fabien Pinaud$^{1,3,4}$}
\email{pinaud@usc.edu}
\author{Christoph A. Haselwandter$^{1,2}$}
\email{cah77@usc.edu}
\affiliation{$^1$Department of Physics and Astronomy, $^2$Department of Quantitative and Computational Biology, $^3$Department of Biological Sciences, and $^4$Department of Chemistry, University of Southern California, Los Angeles, CA 90089, USA}
\date{\today}


\maketitle

\section{Measurement of $\Delta 95$-$99$ diffusion coefficients under mechanical stress and cluster maps}
\label{secNewData}

\begin{figure*}[t!]
	\center
	\includegraphics[width=\textwidth]{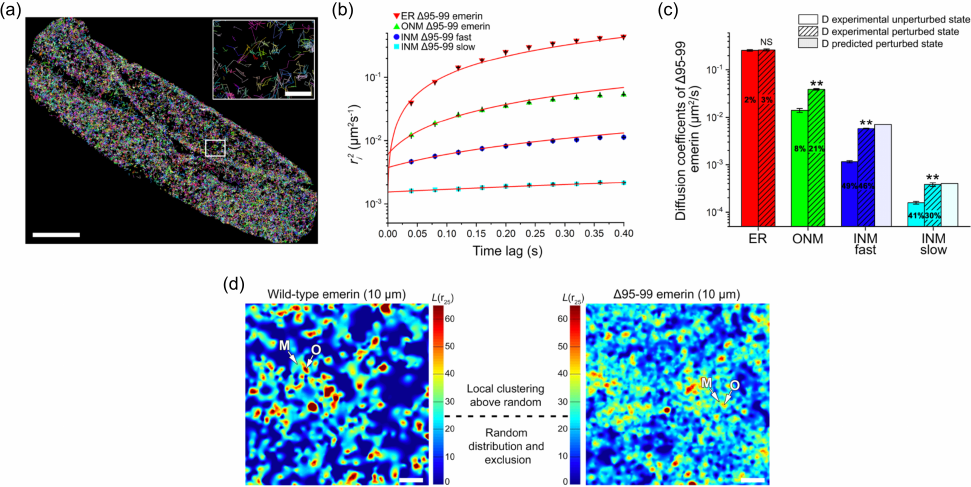}
	\caption{Emerin $\Delta 95$-$99$ mutant displays increased mobilities in response to nuclear mechanical stress at levels predicted by our reaction-diffusion model [see Eqs.~(1)--(4) in the main text]. (a) Trajectory map of individual $\Delta 95$-$99$ emerin at the nuclear envelope in a cell subjected to nuclear mechanical stress on a 10~$\mu$m wide micropattern. The crumpling of the nuclear envelope characteristic of this mutated form of emerin is visible. Scale bar, 5~$\mu$m. Inset: zoom on individual diffusion trajectories of $\Delta 95$-$99$ emerin from the region of interest. Scale bar, 500~nm. (b) Square displacement curves ($r^2_i$) of $\Delta 95$-$99$ emerin diffusive behaviors at the endoplasmic reticulum (ER) and outer (ONM) and inner (INM) nuclear membranes fitted over the first four values with a Brownian diffusion model (red line). (c) Experimental diffusion coefficients ($D$, $\pm$ s.e.m.) and population percentages for $\Delta 95$-$99$ emerin in non-patterned cells (unperturbed, 76944 trajectories, 14 nuclei) or after nuclear deformation on 10~$\mu$m wide micropatterns (perturbed, 73336 trajectories, 13 nuclei) compared to the predicted diffusion coefficients of INM emerin in the perturbed state. T-test comparing experimental unperturbed and perturbed states, NS: non-significant, $^\star$$^\star$: $p<0.01$. (d) Local cluster maps of WT and $\Delta 95$-$99$ emerin after nuclear deformation and perturbation on 10~$\mu$m wide micropatterns. Cluster search radius is 25~nm $\left[L\left(\mathrm{r}_{25}\right)\right]$. M: ``monomeric'' membrane areas (containing primarily rapidly-diffusing emerin), O: ``oligomeric'' nanodomains (containing primarily slowly-diffusing emerin). Scale bar, 250~nm. Panel (c) is, in part, reproduced from the main text and shown here for completeness.}
	\label{fig:1}
\end{figure*}

Figure~\ref{fig:1} provides experimental data supplementary to that shown in Fig.~2(c) of the main text. The experimental data in Fig.~\ref{fig:1} and in Fig.~2(c) of the main text was obtained as follows.

Single molecule tracking of PA-TagRFP-$\Delta 95$-$99$ emerin mutant by photoactivated localization microscopy (sptPALM) at the nuclear envelope of human dermal fibroblasts (HDF) was performed as previously described \cite{Fernandez2022}. In brief, a pEGFP-N1 plasmid backbone encoding PA-TagRFP-$\Delta 95$-$99$ was transfected in HDF stably knocked-down for endogenous emerin by shRNA using X-tremeGENE 360 (Roche) reagent. After 48h transfection, cells were trypsinized, and plated on hexamethyldisilazane-activated glass coverslips (Marienfeld, \#$1.5$, $\cancel{\mathrm{O}} 25$ mm) stamped with $210 \times 10$~$\mu$m fibronectin-micropatterns and blocked with a $1\%$ solution of Pluronic F-127. Cells were allowed to spread on the micropatterns for 6 hours, after which they were briefly washed in $37^{\circ}$C Hank’s Balanced Salt Solution (HBSS) before sptPALM imaging at $37^{\circ}$C in HBSS.

Microscopy imaging was done by highly inclined and laminated optical sheet (HILO) microscopy at the bottom nuclear membrane of cells, on an inverted Nikon Eclipse Ti-E microscope equipped with a 100$\times$/1.49 NA objective (Nikon), an iXon EMCCD camera (Andor), laser lines at 405 and 561 nm (Nikon), a  multiband pass ZET405/488/561/647$\times$ excitation filter (Chroma), a quad-band ZT405/488/561/647 dichroic mirror (Chroma) and a 600/50~nm emission filter (Chroma). Images were acquired continuously at a frame rate of 40~ms per frame, using continuous photoactivation and excitation of PA-TagRFP at 405~nm and 561~nm, and for no longer than 3 minutes per cell to limit UV damage.

Localization by 2D-gaussian fitting of the point spread function of each activated PA-TagRFP-$\Delta 95$-$99$ emerin emitter in each frame, trajectory linkage and tracking analyses were performed using the software package SLIMfast in Matlab as previously described \cite{Fernandez2022}. Trajectories with fewer than three steps were discarded, and diffusion coefficients were estimated using a probability density of square displacement (PDSD) analysis. For each time lag $t$, the PDSD curve was fitted with the following model: 
\begin{equation}
	P\left(\mathbf{r}^2,t\right) = 1- \sum_{i=1}^n a_i(t) e^{-\mathbf{r}^2/\mathbf{r}_i^2(t)}\,, \quad \sum_{i=1}^n a_i(t) =1\,,
\end{equation}
where $\mathbf{r}_i^2(t)$ is the square displacement and $a_i(t)$ is the population density of $i$ numbers of diffusive behaviors at each time lag $t$. Since PA-TagRFP-$\Delta 95$-$99$ emerin at the nuclear envelope of non-micropatterned cells was shown to have four diffusive behaviors \cite{Fernandez2022}, we kept $i = 4$ for PDSD curve fitting of PA-TagRFP-$\Delta 95$-$99$ emerin on 10 $\mu$m wide micropatterns. Square displacement curves $\left(\mathbf{r}_i^2(t)\right)$ were extracted from PDSD analyses and reported with error bars determined using $\frac{\mathbf{r}_i^2(t)}{\sqrt{N}}$, where $N$ is the number of analyzed trajectories per time lag.

The diffusion coefficients ($D$) representative of each of the four PA-TagRFP-$\Delta 95$-$99$ emerin subpopulations were determined by fitting each $\mathbf{r}_i^2(t)$ curves over the first four time lags using OriginPro2022 software and a 2D Brownian diffusion model with position error,
\begin{equation}
	\mathbf{r}^2 = 4 D t + 4 \sigma^2\,.
\end{equation}
All diffusion coefficients $D$ are reported in $\mu$m$^2$s$^{-1}$ $\pm$ standard error of fit value ($\pm$ s.e.m.). Statistical comparisons between $D$ values were done using two-tailed unpaired $t$ tests. Population percentages are derived from the averaged $a_i(t)$ values over the considered time lags. The number of trajectories and nuclei analyzed for PA-TagRFP-$\Delta 95$-$99$ emerin in 10 $\mu$m wide micropatterns was 73336 trajectories in 13 nuclei, similar to those previously obtained for PA-TagRFP-$\Delta 95$-$99$ emerin in cells that were not micropatterned \cite{Fernandez2022} (76994 trajectories, 14 nuclei).

Custer maps of WT and $\Delta 95$-$99$ emerin on 10 $\mu$m wide micropatterns [Fig.~\ref{fig:1}(d)] were produced from superresolution imaging data as previously described \cite{Fernandez2022}, using the Getis and Franklin $L$ function \cite{getis87} and for a distance of 25 nm. In those maps, values $L(\mathrm{r}25)=25$ represent areas where emerin is randomly distributed and values $L(\mathrm{r}25)=70$ represent areas with emerin local density $(70/25)^2 \approx 8$-fold higher than expected for a random distribution.

\section{Estimating emerin concentrations at the nuclear envelope}
\label{secCounts}

\begin{figure*}[t!]
	\center
	\includegraphics[width=0.66\textwidth]{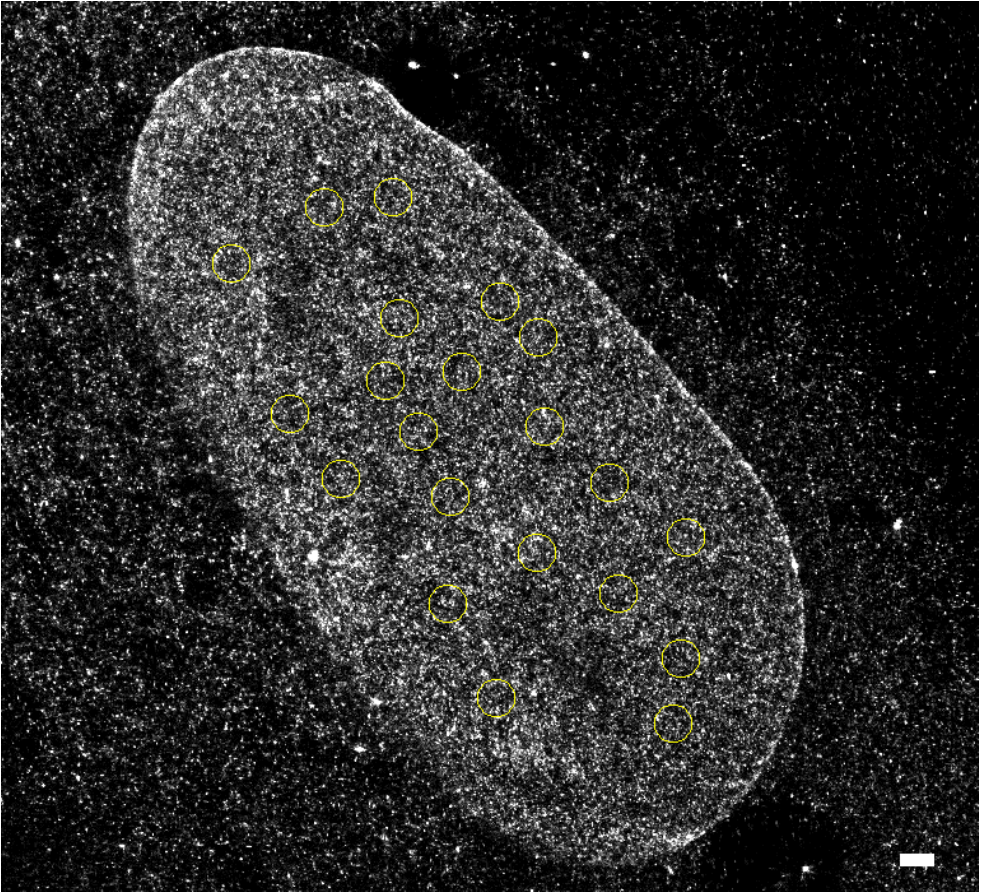}
	\caption{Emerin concentration at the nuclear envelope. The $21$~yellow circles are ROIs, each about $1070$~nm in diameter, in which the total number of emerin were counted with minor corrections to limit over-counting. Scale bar, $1$~$\mu$m.}
	\label{fig:2}
\end{figure*}

In this section we provide a simple estimate of the area number density of emerin at the nuclear envelope. On this basis, we then estimate the fraction of the nanodomain area covered by emerin. In Fig.~\ref{fig:2} we provide an example of an image showing the localization of WT emerin in cells under no mechanical stress. The 21 yellow circles are regions of interest (ROIs) where we counted the total number of emerin detections, after minor corrections to limit overcounting. For simplicity, we used here cells under conditions that do not induce emerin clustering (SUN1 siRNA), so as to allow a global understanding of emerin density at the nuclear envelope without having to separate clustered and ``free'' emerin. From Fig.~\ref{fig:2} we find that there are $3648\pm429$ emerin per $898746.5$~nm$^2$ of nuclear envelope (we used 21 circular ROIs in Fig.~\ref{fig:2}, each about 1070~nm in diameter), yielding about $4000\pm500$~emerin per~$\mu$m$^2$ of~NE.

Based on the above estimates, we can calculate the fraction of the nanodomain area covered by WT emerin under no mechanical stress [see Fig.~1(a) of the main text]. In this case, each nanodomain is found experimentally to contain about eight times more emerin than random. Given that a random distribution of emerin at the nuclear envelope has a density of about $4000\pm500$~emerin per $\mu$m$^2$ (see Fig.~\ref{fig:2}), that approximately $90\%$ of the emerin measured at the nuclear envelope is associated with the INM \cite{Fernandez2022}, and that, roughly, a single emerin molecule occupies an area $1$~nm$^2$--$4$~nm$^2$ at nuclear membranes \cite{Eskandari1998}, this brings the density of emerin inside nanodomains to about 8--10~emerin per nandomain for the 20~nm diameter emerin nanodomains in Fig.~1(a). We thus find that about $3$--$10\%$ of the nanodomain area is covered by emerin, yielding $c\approx0.03$--$0.1$ for WT emerin under no mechanical stress (see the main text).

\section{Emerin diffusion in heterogeneous media}
\label{secDiffModel}

As discussed in the main text, experiments show that WT emerin cluster at the INM to form stable nanodomains that tend to coincide with regions in diffusion maps with slowed-down emerin diffusion (diffusion coefficient $\nu_\mathrm{slow}\approx3\times10^{-4}\mu\mathrm{m}^2/\mathrm{s}$ for WT emerin), while membrane regions outside emerin nanodomains tend to show more rapid emerin diffusion (diffusion coefficient $\nu_\mathrm{fast}\approx2\times10^{-3}\mu\mathrm{m}^2/\mathrm{s}$ for WT emerin) \cite{Fernandez2022}. We also note that about $90\%$ of the WT emerin population at the nuclear envelope is found to be localized to the INM with about $50\%$ of the emerin population at the nuclear envelope being associated with nanodomains, which means that about $56\%$ of the INM emerin are associated with nanodomains. Moreover, local cluster maps of emerin revealed that WT emerin nanodomains roughly cover $20\%$ of the INM area and that the relative density of WT emerin inside nanodomains was increased about eight-fold compared to a random emerin distribution. We show here that, treating the INM as a two-dimensional heterogeneous medium with the observed differences in WT emerin diffusion coefficients, one finds at steady state a fraction of emerin molecules inside nanodomains, $N_\mathrm{slow}$, and a relative density of emerin inside nanodomains, $\langle \sigma_\mathrm{slow}\rangle/\langle \sigma_\mathrm{fast}\rangle$, that roughly agree with experiments on WT emerin. As pointed out in the main text, these results suggest that the observed distributions of emerin along the INM can be understood quantitatively from emerin's diffusion properties, which we take as our starting point for the reaction-diffusion model of emerin nanodomain self-assembly described in the main~text.

Starting with the (stochastic) master equation (ME) for diffusion in heterogeneous media, we follow here Ref.~\cite{Li2017Sep} to obtain exact analytic solutions for the steady-state distribution of emerin. We assume, for now, that steric constraints arising from the finite size of emerin can be neglected, an assumption that can easily be lifted \cite{Li2017Sep}. Solutions of the ME at steady state correspond to zero net emerin fluxes across the nanodomain boundaries, and yield a uniform (average) distribution of emerin inside and outside nanodomains. In particular, analytic solution of the ME shows that, in the steady state of the system, the fraction of emerin inside nanodomains is given by \cite{Li2017Sep}
\begin{align}
	N_\mathrm{slow} = \left(1+\frac{\Gamma_\mathrm{fast}}{\Gamma_\mathrm{slow}}\right)^{-1} \,,
	\label{eq:sol1}
\end{align}
where $\Gamma_\mathrm{slow}$ and $\Gamma_\mathrm{fast}$ are the characteristic times randomly diffusing emerin molecules spend inside and outside emerin nanodomains, respectively, with $\Gamma= A/\nu$, where $A$ is the area of the membrane region characterized by the diffusion coefficient $\nu$. We therefore have
\begin{equation} \label{eq:sol1gggg}
N_\mathrm{slow}=\left(1+\frac{A_\mathrm{fast}/A_\mathrm{slow}}{\nu_\mathrm{fast}/\nu_\mathrm{slow}}\right)^{-1}
\end{equation} 
for emerin at the INM, where $A_\mathrm{slow}$ and $A_\mathrm{fast}$ are the total INM areas occupied and not occupied by emerin nanodomains, respectively. Noting from experiments \cite{Fernandez2022} that emerin nanodomains cover about $20\%$ of the available INM area and that the diffusion coefficients are $\nu_\mathrm{slow}=3\times10^{-4}$~$\mu\mathrm{m}^2/\mathrm{s}$ and $\nu_\mathrm{fast}=2\times10^{-3}$~$\mu\mathrm{m}^2/\mathrm{s}$ for WT emerin with no applied forces, Eq.~(\ref{eq:sol1}) yields $N_\mathrm{slow}\approx63\%$ for the steady-state fraction of WT emerin inside nanodomains at the INM with no applied forces. This theoretical estimate aligns quite well with the corresponding experimental estimate $N_\mathrm{slow}\approx56\%$~\cite{Fernandez2022}.

The density of emerin inside and outside nanodomains is given $\langle\sigma_\mathrm{slow}\rangle =N_\mathrm{slow} M/A_\mathrm{slow}$ and $\langle\sigma_\mathrm{fast}\rangle =N_\mathrm{fast} M/A_\mathrm{fast}$, respectively, where $M$ is the total number of emerin molecules at the INM and $N_\mathrm{fast}= 1-N_\mathrm{slow}$. The relative density of emerin inside nanodomains is therefore given by
\begin{align}
	\frac{\langle \sigma_\mathrm{slow}\rangle}{\langle \sigma_\mathrm{fast}\rangle}=\frac{N_\mathrm{slow}}{N_\mathrm{fast}}\frac{A_\mathrm{fast}}{A_\mathrm{slow}} \,.
	\label{eq:sol2}
\end{align}
Substitution of Eq.~(\ref{eq:sol1gggg}) into Eq.~(\ref{eq:sol2}) results in $\langle\sigma_\mathrm{slow}\rangle/\langle\sigma_\mathrm{fast}\rangle=\nu_\mathrm{fast}/\nu_\mathrm{slow}$. Thus, the ME for diffusion in heterogeneous media predicts that the relative emerin density inside nanodomains is governed by the ratio of the emerin diffusion coefficients outside and inside emerin nanodomains~\cite{Li2017Sep}. For WT emerin, we thus predict $\langle\sigma_\mathrm{slow}\rangle/\langle\sigma_\mathrm{fast}\rangle \approx 7$. Again, this prediction aligns quite well with the corresponding value $\langle\sigma_\mathrm{slow}\rangle/\langle\sigma_\mathrm{fast}\rangle \approx 8$ found experimentally at the INM \cite{Fernandez2022}.

Note that Eq.~(\ref{eq:sol1}) suggests that the steady-state fraction of emerin molecules concentrated inside nanodomains only depends on the fraction of available INM area covered by emerin nanodomains and on the relative emerin diffusion coefficients inside and outside nanodomains, and is independent of the detailed arrangement and shape of emerin nanodomains \cite{Li2017Sep}. We also note that steric constraints arising from the finite size of emerin molecules would somewhat decrease the values of $N_\mathrm{slow}$ and  $\langle\sigma_\mathrm{slow}\rangle/\langle\sigma_\mathrm{fast} \rangle$ estimated above \cite{Li2017Sep}. While the simple model of emerin diffusion in heterogeneous media considered above is able to account for the observed localization of WT emerin to nanodomains under no applied forces, it tends to be less successful for the nuclei under force or the mutated forms of emerin considered in the main text. Also, by construction, the model considered here is unable to predict the self-assembly or size of emerin nandomains, which we address through the reaction-diffusion model described in the main text.

\section{Linear stability analysis and numerical solution procedure}
\label{secAnalysis}

This section provides supplemental information on the physical model of emerin nanodomains in Eqs.~(1)--(4) of the main text,
\begin{equation}\label{eqIflow}
	\frac{\partial I}{\partial t} = \mathrm{F}\left(I,A\right)+\nu_I\left[\left(1-A\right)\nabla^2I+I\nabla^2A\right]
\end{equation}
and
\begin{equation}\label{eqAflow}
	\frac{\partial A}{\partial t} =\mathrm{G}\left(I,A\right)+\nu_A\left[\left(1-I\right)\nabla^2A+A\nabla^2I\right]\,,
\end{equation}
where the polynomials $\mathrm{F}(I,A)$ and $\mathrm{G}(I,A)$ describe the reaction dynamics of $I$ and $A$ complexes. As discussed in the main text, we impose here the constraint $0\leq I+A\leq1$, which accounts for the finite size of $I$ and $A$ complexes, on all reaction and diffusion processes \cite{Haselwandter2011,Haselwandter2015,Kahraman2016,Li2017May}. This constraint produces the non-linear modifications to the standard diffusion terms $\nu_I\nabla^2I$ and $\nu_A\nabla^2A$ in Eqs.~(\ref{eqIflow}) and (\ref{eqAflow}).

For our model to support the self-assembly of emerin nanodomains via a Turing mechanism \cite{Turing1952,Epstein1998,Haselwandter2011,Haselwandter2015,Kahraman2016,Li2017May}, Eqs.~(\ref{eqIflow}) and~(\ref{eqAflow}) must exhibit a non-trivial, stable homogeneous fixed point, $\left(I,A\right)= \left(\bar I, \bar A\right)$,
\begin{equation}\label{eqZeroFlow}
	\mathrm{F}\left(\bar I,\bar A\right) = 0,\hspace{1cm}	\mathrm{G}\left(\bar I,\bar A\right) = 0\,,
\end{equation}
with $\bar I \neq 0,1$ and $\bar A \neq 0,1$. In the absence of diffusion, random perturbations of this steady state must decay over time, yielding the conditions \cite{Haselwandter2011,Haselwandter2015}
\begin{equation}\label{eqTrJ}
	\mathrm{Tr}[\mathrm{\mathbf{\bar J}}]=I_{11}+A_{22}<0
\end{equation}
and
\begin{equation}\label{eqDetJ}
	\mathrm{Det}[\mathrm{\mathbf{\bar J}}]=I_{11}A_{22} - I_{12}A_{21}>0\,,
\end{equation}
where the linear stability matrix
\begin{align}\label{eqJ}
	\mathrm{\mathbf{\bar J}}=
	\begin{pmatrix}
		I_{11} & I_{12} \\
		A_{21} & A_{22}
	\end{pmatrix}
	\equiv 
	\begin{pmatrix}
		\frac{\partial\mathrm{F}}{\partial I}\big\rvert_{\left(I,A\right) = \left(\bar I,\bar A\right)} & \frac{\partial\mathrm{F}}{\partial A}\big\rvert_{\left(I,A\right) = \left(\bar I,\bar A\right)} \\
		\frac{\partial\mathrm{G}}{\partial I}\big\rvert_{\left(I,A\right) = \left(\bar I,\bar A\right)} & \frac{\partial\mathrm{G}}{\partial A}\big\rvert_{\left(I,A\right) = \left(\bar I,\bar A\right)}
	\end{pmatrix}\,.
\end{align}
Equations~(\ref{eqTrJ}) and~(\ref{eqDetJ}) ensure the stability of the reaction-only system under (small) perturbations.

Next, we allow for diffusion of $I$ and $A$ complexes, which yields the joint conditions on the reaction-diffusion processes in Eqs.~(1)--(4) of the main text that must be satisfied to achieve a Turing instability. Allowing for random perturbations of the $I$ and $A$ concentration fields about $I = \bar I$ and $A = \bar A$, it follows from Eqs.~(\ref{eqIflow})--(\ref{eqZeroFlow}) that we must have \cite{Haselwandter2011,Haselwandter2015}
\begin{align}\label{eqRealkappa}
	\left\{\nu_I\left[A_{22}\left(1-\bar A\right)-A_{21}\bar I\right] + \nu_A\left[I_{11}\left(1-\bar I\right)-I_{12}\bar A\right]\right\}^2 - 4\nu_I\nu_A\left(1-\bar I-\bar A\right)\left(A_{22}I_{11}-A_{21}I_{12}\right)>0
\end{align}
and
\begin{equation}\label{eqNegDetJminD}
	\nu_I\left[A_{22}\left(1-\bar A\right)-A_{21}\bar I\right] + \nu_A\left[I_{11}\left(1-\bar I\right)-I_{12}\bar A\right] >0
\end{equation}
for a Turing instability to occur in our model of emerin nanodomain self-assembly. Finally, we note that an estimate of the characteristic length scale arising from a Turing instability, $\ell_c$, can be obtained from the midpoint of the band of unstable perturbation modes \cite{Haselwandter2011,Haselwandter2015},
\begin{align}\label{eqellc}
		\ell_c = \sqrt{\frac{8\pi^2\nu_I\nu_A(1-\bar I-\bar A)}{\nu_I[A_{22}(1-\bar A)-A_{21}\bar I] + \nu_A[I_{11}(1-\bar I)-I_{12}\bar A]}}\,.
\end{align}

\subsection{Reaction kinetics of emerin nanodomains}
\label{secKinetics}

As explained in the main text, we consider here the reaction dynamics
\begin{equation}\label{eqF}
	\mathrm{F}\left(I,A\right)= -f_1I + f_2SA
\end{equation}
and
\begin{equation}\label{eqG}
	\mathrm{G}\left(I,A\right)= -g_1A + g_2SA + \frac{g_3}{2\bar A}SA^2
\end{equation}
of $I$ and $A$ complexes, which yield the homogeneous fixed point
\begin{equation}\label{eqbarAI}
	\bar I =\frac{\left(1-\hat{g}\right)\hat{g}f_2}{f_1+\hat{g}f_2}
	\,,\hspace{1cm}	\bar A = \frac{\left(1-\hat{g}\right)f_1}{f_1+\hat{g}f_2}\,,\hspace{1cm} \hat{g}=\frac{2g_1}{2g_2+g_3} \,.
\end{equation}

Our reaction-diffusion model of emerin nanodomain self-assembly assumes that $I$ complexes act as inhibitors and $A$ complexes as activators of increased $I$ and $A$ concentrations at the INM, leading to $I_{11}<0$ and $A_{22}>0$ in Eq.~(\ref{eqJ}), respectively. Equations~(\ref{eqJ}) and~(\ref{eqF}) thus yield
\begin{equation}\label{eqI11}
	I_{11}=-f_1-f_2\bar A <0\,,
\end{equation}
while Eqs.~(\ref{eqJ}) and~(\ref{eqG}) yield
\begin{equation}\label{eqA22}
	A_{22} = -g_1 + g_2\left(1 - \bar I - 2\bar A\right) + g_3\left(1-\bar I-\frac{3}{2}\bar A\right) >0\,.
\end{equation}

Note from  Eq.~(\ref{eqDetJ}) that we must have $I_{11}A_{22} - I_{12}A_{21} >0$ for a Turing instability to occur. Since $I_{11} < 0$ and $A_{22} >0$, this means that $I_{12}$ and $A_{21}$ must have opposite signs. In our reaction-diffusion model of emerin nanodomain self-assembly, we do not have any reactions in which $I$ complexes stabilize $A$ complexes. Equations~(\ref{eqJ}) and~(\ref{eqG}) therefore mean that
\begin{equation}\label{eqA21}
	A_{21}= -\left(g_2 +\frac{g_3}{2}\right)\bar A < 0\,.
\end{equation}
Equations~(\ref{eqJ}) and~(\ref{eqF}) then yield
\begin{equation}\label{eqI12}
	I_{12} = f_2\left(1-\bar I-2\bar A\right) > 0\,.
\end{equation}
For emerin nanodomain self-assembly to occur through a Turing instability in Eqs.~(1)--(4) of the main text, the reaction rates $f_1$, $f_2$, $g_1$, $g_2$, and $g_3$ in Eqs.~(\ref{eqF}) and~(\ref{eqG}) must satisfy for given, measured $\nu_I$ and $\nu_A$ the constraints in Eqs.~(\ref{eqTrJ}), (\ref{eqDetJ}), (\ref{eqRealkappa}), (\ref{eqNegDetJminD}), and (\ref{eqI11})--(\ref{eqI12}).

\subsection{Numerical implementation}
\label{secNum}

The reaction and diffusion processes considered in our model imply, via the Turing mechanism for nonequilibrium pattern formation, the characteristic length scale $\ell_c$ in Eq.~(\ref{eqellc}), which we corroborate through numerical solutions of our reaction-diffusion equations. The corresponding characteristic diameter of emerin nanodomains, $\ell$, is given by $\ell \approx \ell_c/2$. The characteristic time scale for the self-assembly of emerin nanodomains, $\tau$, can be estimated as $\tau \approx \ell_c^2/\nu_A$, where $\nu_A < \nu_I$ is the diffusion coefficient of the slowly-diffusing particle species in our model (i.e., the $A$ complexes). For our numerical solutions of Eqs.~(1) and~(2) in the main text we employed the \textit{DifferentialEquations} library in Julia \cite{Bezanson2017,Rackauckas2017}. We compared numerical solutions obtained with a range of solvers implemented in this library---including \textit{BS3}, \textit{Tsit5}, \textit{Runge-Kutta}, and \textit{GMRES}---and found similar results. We obtained the numerical solutions in the main text using the \textit{GMRES} solver. We used periodic boundary conditions with a system size $400\times400$~$\mathrm{nm}$$^2$, which is significantly larger than the size of emerin nanodomains observed in experiments \cite{Fernandez2022}. Smaller system sizes approaching the size of emerin nanodomains can yield finite-size artifacts in the emerin patterns generated by our model. We used a $150\times150$ grid for our numerical solutions, and checked that a finer grid produced similar results.

\begin{figure}[t!]
	\includegraphics[width=0.64\textwidth]{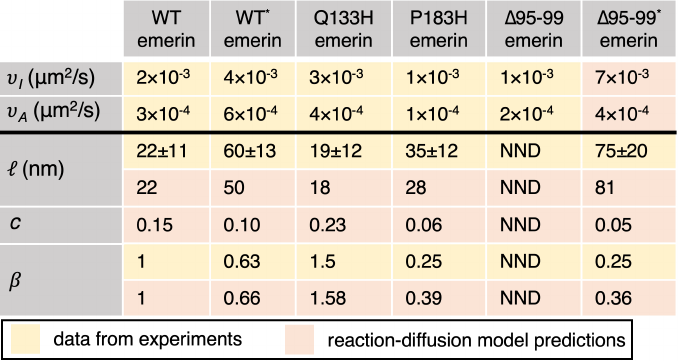}
	\centering
	\caption{Summary of predictions of our reaction-diffusion model (red) and corresponding experimental results (orange) for WT, WT$^\star$,Q133H, P183H, $\Delta95$-$99$, and $\Delta95$-$99^\star$ emerin. NND: No nanodomains.}
	\label{fig:4}
\end{figure}

Figure~\ref{fig:4} summarizes the values of $\nu_I$ and $\nu_A$ for each scenario considered in the main text, as well as the corresponding values of $\ell$ and $c$ obtained from our reaction-diffusion model and measured experimentally. The numerical solutions in Figs.~1 and~2(a,b) of the main text were obtained from initial conditions that were perturbed randomly about the homogeneous steady state, $\left(I,A\right)=\left(\bar{I},\bar{A}\right)$, with the random perturbations being drawn from a uniform distribution over $[-0.0005, 0.0005]$. For each scenario considered in the main text, the reaction rates in our model fix the values of $\bar I$ and $\bar A$ and satisfy the constraints for a Turing instability in Eqs.~(\ref{eqTrJ}), (\ref{eqDetJ}), (\ref{eqRealkappa}), (\ref{eqNegDetJminD}), and (\ref{eqI11})--(\ref{eqI12}), with the exception of our solutions for the $\Delta95$-$99$ emerin system under no mechanical stress, which does not yield a Turing pattern.

\section{Crowding effects arising from a background pool of emerin}
\label{secMonomerCrowd}

Our reaction-diffusion model of emerin nanodomain self-assembly assumes that $I$ and $A$ complexes can both assemble from or dissociate into a pool of emerin and NBPs that lack the molecular requirements to form $I$ or $A$ complexes. To test to what extent steric constraints arising from such a ``background'' pool of emerin (or other NBPs) might affect emerin nanodomain self-assembly, we allowed in our model of $I$ and $A$ reaction and diffusion processes for a modified steric repulsion term $S=\left(1-I-A-m\right)$, where $m$ is the fractional INM area occupied by this background emerin concentration. To estimate~$m$, we note that there are about $4000$~emerin per $\mu$m$^2$ of nuclear envelope (see Sec.~\ref{secCounts}) and that about $40\%$ of the emerin population at the nuclear envelope are observed to be diffusing rapidly at the INM. Assuming that a single emerin molecule occupies an area of about $2$~nm$^2$ at nuclear membranes~\cite{Eskandari1998} and that about half of the rapidly-diffusing emerin at the INM belong to the pool of emerin lacking the molecular requirements to form $I$ or $A$ complexes, we thus have $m\approx 0.2\%$. Inserting $m=0.002$ into our model for WT emerin under no mechanical stress and leaving all other model parameters unchanged we found emerin nanodomains that were about $10\%$ smaller in diameter and with $\beta\approx1.2$ compared to the results in Fig.~1(a) of the main text. Using values of $m$ smaller or not much greater than $m=0.002$ also yielded similar results as in Fig.~1(a) of the main text. We tested whether additional steric effects arising from a pool of emerin (or other NBPs) not accounted for through $I$ and $A$ complexes could effectively be compensated in our reaction-diffusion model by decreasing the assembly rate of $I$ (``crowder'') complexes. Indeed, setting $m=0.002$ in our model for WT emerin under no mechanical stress and decreasing $f_2$ by $20\%$ yielded results that were nearly identical to those in Fig.~1(a) of the main text.

\section{Spontaneous self-assembly of $I$ and $A$ complexes}
\label{secSpontDimer}

To test to what extent our model predictions change if one allows for the spontaneous formation of $I$ or $A$ complexes, we extended our reaction scheme in Eqs.~(3) and~(4) of the main text to allow for the reactions $\varnothing \xrightarrow{f_0} I$ and $\varnothing \xrightarrow{g_0} A$ with the reaction rates $f_0$ and $g_0$, respectively. We thus modified Eqs.~(3) and~(4) of the main text so that these expressions take the form
\begin{equation}\label{eqFspont}
	\mathrm{F}\left(I,A\right) = f_0S\bar I-f_1 I+f_2SA
\end{equation}
and
\begin{equation}\label{eqGspont}
	\mathrm{G}\left(I,A\right) = g_0S\bar A-g_1 A+g_2SA + \frac{g_3}{2 \bar A} S A^2\,,
\end{equation}
respectively, with
\begin{equation}\label{eqbarAIdimer}
	\bar A = \frac{\left(1-\hat{g}\right)\left(f_1-\hat{g}f_0\right)}{f_1+\hat{g}\left(f_2-f_0\right)}\,,\hspace{1cm}	\bar I =\frac{\left(1-\hat{g}\right)\hat{g}f_2}{f_1+\hat{g}\left(f_2-f_0\right)}\,,\hspace{1cm} \hat{g}=\frac{2g_1}{2 \left(g_0+g_2\right) +g_3} \,.
\end{equation}
For the WT and mutant emerin systems with and without applied forces considered in the main text, we found that including finite $f_0 \lesssim 0.1 f_1$ and $g_0 \lesssim 0.1 f_1$, while decreasing $f_2$ and $g_2$ slightly so as to compensate for the increased assembly of $I$ and $A$ complexes, produced results similar to those in Figs.~1 and~2(a,b) of the main text.

\bibliographystyle{apsrev4-2}

\bibliography{paper}